\documentclass[12pt,english]{article}
\usepackage[T1]{fontenc}
\usepackage{amsmath}
\usepackage{amssymb}
\usepackage{graphicx}
\usepackage{subcaption}
\usepackage{booktabs} 
\usepackage{array}    
\usepackage{tabularx} 

\usepackage{graphicx}
\graphicspath{ {images/} }

\makeatletter
\usepackage{jheppub}
\def\@fpheader{\relax}
\allowdisplaybreaks[1]

\makeatother

\usepackage{babel}

\subheader{}

\title{Semiclassical resolution of the black hole singularity inspired in the minimal uncertainty approach}
\author[a]{Brayan Melchor,}
\author[a]{Rolando Perca,}
\author[b]{Wilfredo Yupanqui.}

\affiliation[a]{Escuela Profesional de F\'isica, Universidad Nacional de San Agust\'in de Arequipa,\protect\\ Av. Independencia, Arequipa, Arequipa, Per\'u}  
\affiliation[b]{Departamento de F\'isica, Divisi\'on de Ciencias e Ingenier\'ias, Universidad de Guanajuato, Loma del Bosque 103, Le\'on 37150, Guanajuato, M\'exico.}

\emailAdd{bmelchor@unsa.edu.pe}
\emailAdd{rperca@unsa.edu.pe}
\emailAdd{w.yupanquicarpio@ugto.mx}

\abstract{We propose a new lapse function that simplifies the Hamiltonian constraint, describing the interior of the black hole in terms of the Ashtekar-Barbero variables, into a more straightforward form. The new Hamiltonian leads to different equations of motion than those found in the literature, but through a suitable transformation between temporal parameters, it is found that such a choice leads us to the classical solutions of the Schwarzschild metric, still preserving the physical singularity. In order to resolve this singularity, and inspired by the minimal uncertainty approach, we modify the classical algebra between the dynamic variables of the model, imposing an effective dynamics within the black hole. As a consequence, one of the dynamic variables, denoted by $p_b$, acquires a minimum value at the singularity $t=0$, and on the other hand, the variable related to the radius of the 2-sphere, $p_c$, leads to the resolution of the classical singularity of the black hole by replacing it with a bounce that connects the interior of the black hole with the interior of the white hole. This bounce occurs in the Planck-scale region, where a new event horizon manifests. Upon crossing this horizon, the nature of the interval changes from spatial to temporal outside the white hole.
}

\begin{document}

\maketitle
\flushbottom

\section{Introduction}
\label{sec:intro}
One of the most important challenges of quantum gravity is the black hole's interior as well as its hidden singularity. It is at the singularity where the predictions of General Relativity meet the paradoxes and limitations of quantum mechanics.

In this paper, we analyze the interior of a Schwarzschild black hole using the Ashtekar-Barbero variables, which consist of the connection and its conjugate momentum, the densitized triads \cite{Abhay_Ashtekar_2004,Thiemann2007, Thiemann2003}. Unlike previous works \cite{Corichi:2015xia}, \cite{Chiou2008}, and \cite{Bosso}, we adopt a different lapse function. Our approach demonstrates that the new Hamiltonian we obtain leads to classical equations of motion, ultimately recovering the Schwarzschild metric.

Instead of employing polymer quantization \cite{Abhay_Ashtekar_2003, PhysRevD.76.044016, PhysRevD.95.065026, PhysRevD.92.104029, FLORESGONZALEZ2013394}-a technique that resolved classical singularities by introducing a bounce from a black hole to a white hole in the vacuum case-we modify the Poisson algebra of Ashtekar variables \cite{Bosso}. This modification is inspired by the minimal uncertainty approach, a quantum gravity model that revises Heisenberg's uncertainty relation to incorporate gravity in quantum measurement processes \cite{PhysRevD.52.1108, G.Veneziano_1986, SCARDIGLI199939, BIZET2023137636}.

The key innovation lies in resolving the black hole singularity, resulting in a bounce occurring at a minimum value of the two-sphere radius. We introduce a new lapse function chosen to facilitate the quantization process and align with established results \cite{10.1088/1361-6382/ad4fd7}.

The paper is organized as follows: In Section \ref{sec:The interior of Schwarzschild black hole}, we provide a brief description of the interior of the Schwarzschild black hole in terms of Ashtekar-Barbero variables. In Section \ref{sec:Non-deformed dynamics of black hole interior}, we discuss classical dynamics within the black hole's interior, considering a different proposal for the lapse function. The section \ref{sec:Minimal_uncertainty_approach} is devoted to discussing the minimal uncertainty approach proposed in different quantum gravity models. In Section \ref{sec:Effective dynamics in the interior of Schwarzschild black hole}, we explore a modification in the Poisson algebra to obtain effective dynamics that resolve the physical singularity of the black hole. Finally, in Section \ref{sec:Conclusions}, we present the conclusions.

\section{The interior of Schwarzschild black hole}
\label{sec:The interior of Schwarzschild black hole}
 From the metric of the Schwarzschild black hole 
\begin{equation}
ds^{2}=-\left(1-\frac{2GM}{r}\right)dt^{2}+\left(1-\frac{2GM}{r}\right)^{-1}dr^{2}+r^{2}d\Omega^2,
\end{equation}
with $r\in(0,\infty)$ as the radial coordinate and the radius of the two spheres in Schwarzschild coordinates $\left(t,r,\theta,\phi\right)$, one can obtain the interior metric by interchanging $t$ and $r$, as follows
\begin{equation}
ds^{2}=-\left(\frac{2GM}{t}-1\right)^{-1}dt^{2}+\left(\frac{2GM}{t}-1\right)dr^{2}+t^{2}d\Omega^2.\label{eq:sch-inter}
\end{equation}
Here, and throughout this work, $t$ is the Schwarzschild time coordinate with a range of $t\in(0,2GM)$ in the interior.

The utilization of Ashtekar variables allows us to obtain the expression of the Hamiltonian constraint that applies to the inner region of Schwarzschild \cite{Ashtekar}. The symmetric reduced Hamiltonian constraint adapted to this model, which governs the dynamics of the black hole interior, is formulated in relation to the Ashtekar-Barbero variables as follows \cite{Chiou2008}
\begin{equation}
H=-\frac{8\pi N}{\gamma^2}\frac{\mathrm{sgn}(p_c)}{\sqrt{|p_{c}|}}\left[2bcp_c+\left(b^{2}+\gamma^{2}\right)p_{b}\right],\label{eq:H-class-N}
\end{equation}
here $N$ represents the lapse function and $\text{sgn}(p_c)$ is the sign function. The Hamiltonian \eqref{eq:H-class-N} is written in terms of two independent components, traditionally
called the connections $b(T),\,c(T)$, and their conjugate momenta are called densitized triads $p_{b}(T)$ and $p_{c}(T)$, with $T$ as a generic time parameter.

The diffeomorphism constraint vanishes identically due to homogeneous
nature of the model \cite{Barbero}, i.e.,
\begin{equation}
\left(b^{2}+\gamma^{2}\right)\frac{p_{b}}{b}+2cp_{c}\approx 0.\label{eq:weak-van}
\end{equation}
Furthermore, the Poisson brackets between the conjugate variables turns out to be
\begin{align}
\{c,p_{c}\}= & 2G\gamma, & \{b,p_{b}\}= & G\gamma.\label{eq:classic-PBs-bc}
\end{align}
The Schwarzschild interior metric in these variables is expressed as
\begin{equation}
\label{eq: Sch in v. Asht}
ds^{2}=-N(T)^{2}dT^{2}+\frac{p_{b}^{2}}{L_{0}^{2}|p_{c}|}dx^{2}+|p_{c}|(d\theta^{2}+\sin^{2}\theta d\phi^{2}),
\end{equation}
with the metric components given by
\begin{align}
g_{xx}\left(T\right)= & \frac{p_{b}\left(T\right)^{2}}{L_{0}^{2}|p_{c}\left(T\right)|},\label{eq:grrT}\\
g_{\theta\theta}\left(T\right)= & \frac{g_{\phi\phi}\left(T\right)}{\sin^{2}\left(\theta\right)}=g_{\Omega\Omega}\left(T\right)=|p_{c}\left(T\right)|,\label{eq:gththT}
\end{align}
where its classical singularity resides at $p_{b}\to0,p_{c}\to0$.

Given the topology
$\mathbb{R}\times\mathbb{S}^{2}$ of the spatial sector of the model,
in order to prevent the integral involved in the computation of the symplectic
structure in $\mathbb{R}$, one needs to restrict the radial part
to an auxiliary length $L_{0}$ which restricts the volume of the
integration to $V_{0}=a_{0}L_{0}$, where $a_{0}$ is the area of
the 2-sphere $\mathbb{S}^{2}$ in $L_{0}\times\mathbb{S}^{2}$ \cite{Ashtekar}.
This is possible due to the homogeneity of the model since one can take
the limit of $L_{0}\to\infty$ at the end. Clearly, none of the physical
results should depend on $L_{0},\,a_{0}$ or their rescalings. The physical volume of $L_{0}\times\mathbb{S}^{2}$ is \cite{Ashtekar, Chiou:2008nm}
\begin{equation}
\label{eq: Vcell}
    \int_{L_{0}\times\mathbb{S}^{2}} \sqrt{|\det E|} d^3x = 4\pi\sqrt{|p_c|}|p_b|.
\end{equation}
The three surfaces on the sphere of interest, each with its distinct geometric characteristics and denoted by $S_{x,\phi}$, $S_{x,\theta}$ and $S_{\theta,\phi}$ within the integration interval of \eqref{eq: Vcell}, have their respective physical areas provided in \cite{Ashtekar} 
\begin{equation}
    A_{x,\theta}=A_{x,\phi}=2\pi p_b, \quad \quad A_{\theta,\phi}=\pi p_c. \label{eq:pb_pc_physical_area}
\end{equation}
This gives the physical meanings of the triad variables $p_b$ and $p_c$.

\section{Non-deformed dynamics of black hole interior}
\label{sec:Non-deformed dynamics of black hole interior}
Choosing an appropriate lapse function in \eqref{eq:H-class-N} allows for the simplification of the Hamiltonian, aiming to make the resulting equations of motion more manageable when solving them. There are several possible choices for $N$, for example, in \cite{Ashtekar, Corichi:2015xia, Modesto}, the following one was considered
\begin{equation}
N\left(T\right)=\frac{\gamma\text{sgn}(p_c)\sqrt{|p_{c}\left(T\right)}|}{16\pi G b}.\label{eq:Lapse_funct_corichi_bosso}
\end{equation}
The reason for this choice is because the equations of motion obtained for the pair $(c,\ p_c)$ decouple from those obtained for the pair $(b,\ p_b)$. In \cite{Bosso}, the same lapse function \eqref{eq:Lapse_funct_corichi_bosso} is considered for a modified algebra inspired by the Generalized Uncertainty Principle (GUP), which allows for the resolution of the black hole singularity. Another choice of $N$ is found in \cite{Chiou:2008nm}, which takes the form
\begin{equation}
N\left(T\right)=p_{b}\left(T\right)\sqrt{|p_{c}\left(T\right)|}.\label{eq:Lapse_Chiou}
\end{equation}
This lapse function is chosen to study the dynamics of the Schwarzschild black hole in the context of loop quantum geometry. In this work, we choose another $N$ that allows us to obtain a simple Hamiltonian for quantization, the choice is
\begin{equation}
N\left(T\right)=\frac{\text{sgn}(p_c)\sqrt{|p_{c}\left(T\right)}|}{16\pi G}, \label{eq:lapsNT}
\end{equation}
and from (\ref{eq:H-class-N}), the Hamiltonian becomes in
\begin{equation}
H=-\frac{1}{2G\gamma^2}\left[\left(b^{2}+\gamma^{2}\right)p_b +2bcp_{c}\right]. \label{eq:H-class-1}
\end{equation}
This Hamiltonian, unlike those obtained from \eqref{eq:Lapse_funct_corichi_bosso} and \eqref{eq:Lapse_Chiou}, is more manageable for quantization since it doesn't include the variable $b$ in the denominator and leads to a first-order differential equation in the connection representation \cite{10.1088/1361-6382/ad4fd7}.

Regarding the physical outcomes, the choice of any lapse function mentioned here leads to the usual classical solution. Therefore, the equations of motion obtained from different choices of $N$ are equivalent. The same holds for our case, as we will see below.

\begin{figure}[htb!] 
\begin{center}
\includegraphics[scale=0.6]{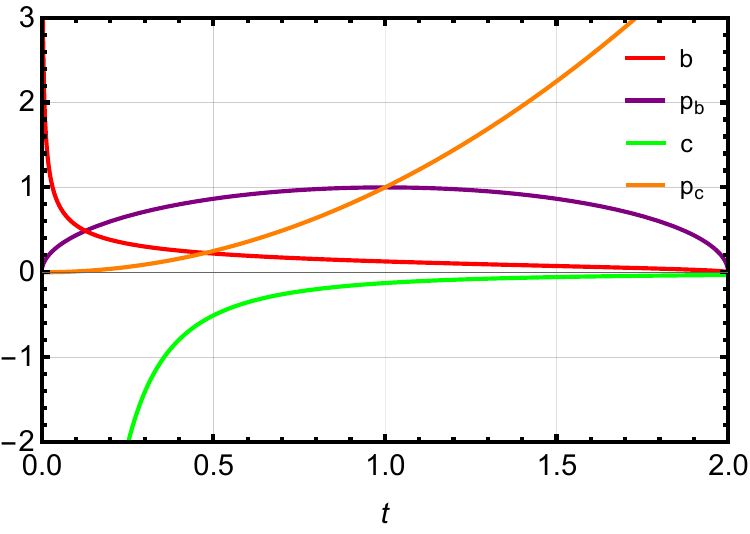} 
\caption{The behavior of canonical variables as a function of Schwarzschild time t. We have chosen the positive sign for $b$ and the negative sign for $c$.}
\label{fig1}
\end{center}
\end{figure}

From the Hamiltonian \eqref{eq:H-class-1} one obtains the equations of motion for the dynamical variables $b(T)$, $p_b(T)$, $c(T)$, and $p_c(T)$, using the classical Poisson algebra \eqref{eq:classic-PBs-bc},
\begin{eqnarray}
    \frac{db(T)}{dT} &=& \left\{ b,H\right\} =-\frac{\left(b^2+\gamma^2\right)}{2\gamma},\label{eq:Class_Equat_motio_b_T}\\
    \frac{dp_b(T)}{dT} &=& \left\{ p_b,H\right\} =\frac{bp_b+c p_c}{\gamma},\label{eq:Class_Equat_motio_pb_T}\\
    \frac{dc(T)}{dT}   &=& \left\{c,H\right\} =-\frac{2bc}{\gamma},\label{eq:Class_Equat_motio_c_T}\\
    \frac{dp_c(T)}{dT} &=& \left\{ p_c,H\right\} = \frac{2bp_c}{\gamma}.\label{eq:Class_Equat_motio_pc_T}
\end{eqnarray}
We can see that the equation for $b$ is decoupled from the other dynamic variables and can be solved through direct integration. The equation for $p_b$ is coupled in terms of these, therefore, to solve it, it must be decoupled first, and this is possible thanks to the constraint \eqref{eq:weak-van}. On the other hand, one can see that the equations of motion for $c$ and $p_c$ are coupled only with $b$, then if we know $b$, is straightforward to find the solutions for them. Then we get the following relations
\begin{eqnarray}
    b(T)&=&\gamma \tan\left(-\frac{T}{2}-C_1\right),\label{eq:Class_Equat_b_T}\\
    p_b(T)&=&C_2 \cos{\left(-\frac{T}{2} -C_1\right)}\sin{\left(-\frac{T}{2} -C_1\right)},\label{eq:Class_Equat_pb_T}\\
    c(T)&=&C_3\sec^4\left(-\frac{T}{2}-C_1\right),\label{eq:Class_Equat_c_T}\\
    p_c(T)&=&C_4\cos^4{\left(-\frac{T}{2} -C_1\right)},\label{eq:Class_Equat_pc_T}
\end{eqnarray}
where $C_i$, for $i=1,\ 2,\ 3,\ 4$, are integration constants. To do practical calculations, we choose $C_1=0$ because it is an arbitrary phase that does not play a significant role for our purposes. To determine the other integration constants, we must express \eqref{eq:Class_Equat_b_T}-\eqref{eq:Class_Equat_pc_T} in terms of the Schwarzschild time $t$. Since $p_c (t)=t^2$ and $p_c=C_4 \cos^4 (-T/2 )$, we identify $t^2=C_4\cos^4 (-T/2 )$. Also, since $t$ has dimensions of the Schwarzschild radius, we set $C_4=4G^2M^2$, the relationship between the Schwarzschild time and the generic parameter $T$ is
\begin{equation}
    t=2GM\cos^2 (-T/2 ). \label{eq:Transf_T_to_t}
\end{equation}
Using this transformation, we can express the remaining variables in terms of $t$, considering the trigonometric identities $\tan{(-T/2)}=\pm\sqrt{\sec^2{(-T/2)}-1}$, $\sin{(-T/2)}=\pm\sqrt{1-\cos^2{(-T/2)}}$, and $\sec{(-T/2)}=1/\cos{(-T/2)}$. The constant $C_2$ is determined using the $g_{xx}(t)$ component of the metric in terms of $t$, as given by \eqref{eq:grrT}, from which we obtain $C_2=2GM L_0$. In relation to the constant $C_3$, this is determined from constraint \eqref{eq:weak-van}, considering that $cp_c = C_3$, and $(b^2 + \gamma^2)p_b/b = \pm 2\gamma G M L_0$. So, we conclude that
\begin{eqnarray}
  b(t)&=&\pm\gamma \sqrt{\frac{2GM}{t}-1},\label{eq:b_usually}\\
  p_b(t)&=&L_0 \sqrt{t(2GM-t)}, \label{eq:b_pb_usually}\\
  c(t)&=&\mp\frac{\gamma GM L_0 }{t^2},\label{eq:c_usually}\\
  p_c(t)&=&t^2.\label{eq:p_c_usually}
\end{eqnarray}
With these results, we conclude that the choice of the lapse function $N(T)$ in \eqref{eq:lapsNT} leads to classical solutions, indicating that such a choice is reasonable. The behavior of these variables in terms of $t$ is shown in Figure \ref{fig1}, for the values $\gamma = \ln{2}/\pi\sqrt{3}$, $L_0 = 1$, and $G = M = 1$, the choice of these values in our paper is purely for representational purposes. From here, it can be observed that both $b$ and $p_b$ are zero at the horizon, and as we approach the region $t = 0$, $b$ grows indefinitely, while $p_b$ approaches zero. On the contrary, both $c$ and $p_c$ have non-zero values at the horizon, but as $t \rightarrow 0$, the variable $c$ decreases towards negative values indefinitely, and $p_c \rightarrow 0$.

At $t=0$, the classical singularity is located. This leads to the divergence of the Kretschmann scalar, which in terms of Ashtekar variables is given by \cite{universe8070349}
\begin{equation}
K=\frac{48 G^2 M^2}{p_c^3}.\label{eq:Kretschmann_scalar}
\end{equation}
On the horizon $t\rightarrow2GM$ and $K=\frac{3}{4G^4M^4}$, which indicates that in this region there is no real singularity. In $t\rightarrow0$ we have that $K\rightarrow\infty$ indicating the presence of a physical singularity there, as expected. From a physical point of view, this means that as we approach the classical singularity, the parameter $p_c$ tends to zero, which implies that the associated quantities, such as Riemann invariants, become infinitely large.

\section{Minimal uncertainty approach}
\label{sec:Minimal_uncertainty_approach}
Heisenberg's Uncertainty Principle (HUP) is a fundamental concept in quantum mechanics that states that certain pairs of physical properties of a particle cannot be precisely measured simultaneously. In those measurement processes, the gravitational interaction between particles was completely neglected, although this was somehow justified by the considerable weakness of gravity when compared with other fundamental interactions. However, gravity should be considered when elementary measurement processes are discussed in order to address fundamental questions in nature. Incorporating gravity into these quantum measurement processes led to the generalization of Heisenberg's uncertainty relation. This modification to the HUP is known as the Generalized Uncertainty Principle (GUP), which introduces additional terms that account for gravitational effects at very small scales, typically at the Planck scale.

Position and momentum are two complementary features of a quantum particle, so they satisfy an uncertainty relation, first derived by H. Kennard \cite{kennard1927quantenmechanik}, in 1927, and H. Weyl \cite{guth1929gruppentheorie}, in 1929, mathematically represented by
\begin{equation}
    \Delta q\Delta p\geq\frac{\hbar}{2},\label{eq:Uncert_relat_q_p_usual}
\end{equation}
where $\Delta q$ and $\Delta p$ represent the standard deviation of the position and the momentum, respectively, and $\hbar$ is the reduced Planck constant. This is a mathematical expression of HUP, which states that it is not possible to simultaneously determine the position and momentum of a quantum particle. The same expression is obtained by considering position and momentum as operators that satisfy the commutation relation
\begin{equation}
    [\hat{q}, \hat{p}] = i\hbar.\label{eq:Conmut_rel_x_p}
\end{equation}
Substituting this into Robertson's uncertainty relation (1929) \cite{PhysRev.34.163}
\begin{eqnarray}
    \Delta A\Delta B\geq\frac{1}{2}\left|\langle\psi|[\hat{A},\hat{B}]|\psi\rangle\right|,\label{enq:Generalz_Heisenbrg_ineqlty}
\end{eqnarray}
we obtain \eqref{eq:Uncert_relat_q_p_usual}.

To introduce the quantum effects of gravity, the uncertainty relation for position and momentum in \eqref{eq:Uncert_relat_q_p_usual} is extended to \cite{PhysRevD.52.1108}
\begin{eqnarray}
    \Delta q\Delta p\geq\frac{\hbar}{2}\left(1+\beta\left(\Delta p\right)^2\right),\label{Eqq.GUP.1}
\end{eqnarray}
where $\beta$ is the deformed parameter. The modified uncertainty relation \eqref{Eqq.GUP.1}, known in the literature as GUP, ensures a non-zero minimal uncertainty in position
\begin{eqnarray}
    \Delta q_0=\hbar\sqrt{\beta},
\end{eqnarray}
and is associated with the quantization of gravity. In \cite{BIZET2023137636}, it is demonstrated that the GUP arises from the consideration of non-extensive entropies. GUP is also derived from different proposals: in \cite{G_Veneziano_1986}, the scattering of strings at ultra-high energies is considered to analyze the divergences of quantum gravity at the Planck scale; in \cite{MAGGIORE199365}, a gedanken experiment is proposed to measure the area of the apparent horizon of black holes in the context of quantum gravity; and \cite{SCARDIGLI199939} explores the idea that spacetime in the Planck region fluctuates, leading to the possibility of virtual micro-black holes affecting the measurement process. On the other hand, in \cite{Bosso_2023} the authors critically discussed some shortcomings and open problems arising within the framework of GUPs, which are often overlooked or naively addressed in the literature.

The modified commutation relation for the $\hat{q}$ and $\hat{p}$ operators associated with (\ref{Eqq.GUP.1}) is expressed as
\begin{eqnarray}
\left[\hat{q},\hat{p}\right]=i\hbar\left(1+\alpha\  \hat{p}^2\right).\label{Eqq.GCR.1}
\end{eqnarray}
Due to the deformed commutator in (\ref{Eqq.GCR.1}) the variables $q$ and $p$ are not conjugates anymore \cite{PhysRevD.52.1108}.

On the other hand, there is another extension to the uncertainty relation \eqref{eq:Uncert_relat_q_p_usual} which, unlike GUP, gives the nonzero minimal uncertainty in the momentum, called Extended Uncertainty Principle (EUP) \cite{PhysRevD.79.125007, MUREIKA201988, doi:10.1142/S0217732319502043}. The mathematical expression for this new minimal uncertainty relation is
\begin{eqnarray}
    \Delta q\Delta p\geq\frac{\hbar}{2}\left(1+\alpha\left(\Delta q\right)^2\right),\label{enq.GUP_q_p_min_p}
\end{eqnarray}
which, as mentioned, leads to a nonzero minimal uncertainty in momentum
\begin{eqnarray}
     \left(\Delta p\right)_{min}=\hbar\sqrt{\alpha}.\label{eq:Min_uncer_momentum}
\end{eqnarray}
The modified commutation relation from which \eqref{enq.GUP_q_p_min_p} follows, via \eqref{enq:Generalz_Heisenbrg_ineqlty}, is expressed as
\begin{equation}
    \left[\hat{q},\hat{p}\right]=i\hbar\left(1+\alpha \hat{q}^2\right).\label{eq:Deformd_conmuta_relat_x_p}
\end{equation}
The corrections from EUP have gained relevance recently, as it was previously not believed to be necessary to introduce quantum corrections on a large scale in gravitational physics. Of course, this is no longer the case, and EUP provides a pathway to introduce quantum effects at macroscopic distances.

So far, we have presented modifications to the Dirac algebra with quadratic corrections as in equations \eqref{Eqq.GCR.1} and \eqref{eq:Deformd_conmuta_relat_x_p}. However, in the literature, there are other modifications with different origins. In particular, in \cite{ALI2009497}, a modification is presented that introduces linear corrections to the Heisenberg algebra
\begin{equation}
   \left[\hat{q}_i,\hat{p}_j\right]=i\hbar\left[\delta_{ij}-\alpha\left(p\delta_{ij}+\frac{p_ip_j}{p}\right)+\alpha^2\left(p^2\delta_{ij}+3p_ip_j\right)\right].\label{eq:DSR_conmut_rel}
\end{equation}
These commutators are consistent with String Theory, Black Holes Physics, and Doubly Special Relativity (DSR). The uncertainty relation associated with \eqref{eq:DSR_conmut_rel}, in 1-dimension, is
\begin{equation}
    \Delta x\Delta p\geq\frac{\hbar}{2}\Biggl[1+\left(\frac{\alpha}{\sqrt{\langle p^2\rangle}}+4\alpha^2\right)\Delta p^2+4\alpha^2\langle p\rangle^2-2\alpha\sqrt{\langle p^2\rangle}\Biggr].
\end{equation}
These in turn imply a minimum measurable length and a maximum measurable momentum
\begin{equation}
    (\Delta x)_{\text{min}}\sim\hbar\alpha,\quad\quad\quad (\Delta p)_{\text{max}}\sim\frac{1}{\alpha}.
\end{equation}

Both \eqref{Eqq.GCR.1}, \eqref{eq:Deformd_conmuta_relat_x_p}, and \eqref{eq:DSR_conmut_rel} serve distinct purposes and consequently lead to different predictions. Particularly, in \cite{Bosso_2020}, the approach of minimum uncertainty in the position is employed in quantum cosmology for a Kantowski-Sachs model describing the interior of a black hole. This new perspective yields a modified Wheeler-DeWitt equation, producing diverse outcomes compared to the conventional approach. Pursuing this notion further, exploring the application of the minimum uncertainty approach to the Kantowski-Sachs model delineated in \eqref{eq: Sch in v. Asht} would be intriguing.

For the standard quantization, we apply the Dirac prescription to the classical algebra \eqref{eq:classic-PBs-bc} to obtain the commutation
        algebra of the canonical variables as
        \begin{align}
         [\hat{b},\hat{p}_{b}] &= i\hbar G\gamma, \label{eq:Usual_Commut_Relation_1}\\ 
         [\hat{c},\hat{p}_c] &= 2i\hbar G\gamma. \label{eq:Usual_Commut_Relation_2}
        \end{align}
As a consequence, in a representation where $p_b$ and $p_c$ act multiplicatively, $b$ and $c$ are given by differential operators
\begin{eqnarray}
    b&= & i\hbar G\gamma\frac{\partial}{\partial p_b},\\ 
    c&= & 2i\hbar G\gamma\frac{\partial}{\partial p_c}.
\end{eqnarray}
The Wheeler-DeWitt equation corresponding to the Hamiltonian \eqref{eq:H-class-N}, with the lapse function \eqref{eq:Lapse_funct_corichi_bosso}, presents serious issues. This is because at $p_c=0$ resides the classical singularity, and in that region, the Wheeler-DeWitt equation is manifestly singular there \cite{Ashtekar}. A similar problem arises when considering the Hamiltonian \eqref{eq:H-class-1}; the presence of the singularity makes it difficult to define a Hilbert space in which to perform the quantization \cite{10.1088/1361-6382/ad4fd7}.

To understand the quantum phenomena occurring within a black hole, one must consider the effects of quantum gravity, namely at Planck scales. Motivated by this, it is natural to contemplate applying the minimum uncertainty approach, which accounts for such effects, to the quantization of the Hamiltonian \eqref{eq:H-class-1} via a Wheeler-DeWitt equation. Considering, from \eqref{eq:Usual_Commut_Relation_1}-\eqref{eq:Usual_Commut_Relation_2}, the corresponding modified commutation relations 
\begin{align}
\left[b,p_{b}\right]= & i\hbar G\gamma\left(1+\beta_{b}b^{2}\right),\label{eq:Mod_com_b_pb}\\
\left[c,p_{c}\right]= & 2i\hbar G\gamma\left(1+\beta_{c}c^{2}\right),\label{eq:Mod_com_c_pc}
\end{align}
one can find the following uncertainty relations 
\begin{align}
\Delta b\Delta p_{b}\geq & \frac{\hbar G\gamma}{2}\left[1+\beta_{b}(\Delta b)^{2}\right],\label{eqn:unc_b_pb}\\
\Delta c\Delta p_{c}\geq & \hbar G\gamma\left[1+\beta_{c}(\Delta c)^{2}\right],\label{eqn:unc_c_pc}
\end{align}
which correspond to minimal uncertainties for $p_{b}$ and $p_{c}$
of the order of $G\gamma\sqrt{\beta_{b}}$ and $2G\gamma\sqrt{\beta_{c}}$,
respectively. However, carrying out the quantization process using this approach is not trivial. Analyzing the structure of the Hilbert space and ensuring that everything is mathematically solid is a crucial step. Due to the extensive analysis required, we will not address it in this work, leaving it for a separate future study.

It is worth noting that another valid modification for \eqref{eq:Usual_Commut_Relation_1}-\eqref{eq:Usual_Commut_Relation_2} could potentially be based on \eqref{eq:DSR_conmut_rel}, incorporating linear corrections. However, in our case, we opted for quadratic corrections due to their greater mathematical simplicity when analyzing the dynamics at hand.

It is expected that an acceptable quantum gravity model resolves, at least effectively, the singularity of black holes. In this regard, in order to test the validity of our proposal, we will study the dynamics inside the black hole, characterized now by the Hamiltonian \eqref{eq:H-class-1}, from a semi-classical (or effective) perspective, by considering a deformed Poisson algebra instead of \eqref{eq:Mod_com_b_pb} and \eqref{eq:Mod_com_c_pc}. That is
\begin{align}
\left\{ b,p_{b}\right\} = & G\gamma\left(1+\beta_{b}b^{2}\right),\label{eqn:b_pb}\\
\left\{ c,p_{c}\right\} = & 2G\gamma\left(1+\beta_{c}c^{2}\right).\label{eqn:c_pc}
\end{align}
Therefore, $\beta_{b}$ and $\beta_{c}$ effectively define the magnitude of the effects introduced with the algebra (\ref{eqn:b_pb})-(\ref{eqn:c_pc}). The specific choices of these parameters enable us to explore the effects on the deformed algebra and analyze their impact on the dynamics and observables of the physical system.

This practice of investigating aspects of classical dynamics using GUP-type deformations is not new in the literature; for example, references \cite{Bosso,PhysRevD.65.125028,PhysRevD.66.026003} can be mentioned. However, reducing the commutation relations \eqref{eq:Mod_com_b_pb}-\eqref{eq:Mod_com_c_pc} to their classical limit \eqref{eqn:b_pb}-\eqref{eqn:c_pc} is more involved than simply applying the transformation $\frac{[\hat{q}_i,\hat{p}_i]}{i\hbar}\to\{q_i,p_i\}$. In \cite{CASADIO2020135558}, it has been shown that it is impossible to estimate any effect of the GUP in the classical limit. On the other hand, \cite{Bosso_2023} discusses the consistent limits of the GUP, showing that, contrary to recent statements in \cite{CASADIO2020135558}, it can be well-motivated to analyze corrections to classical dynamics from Planck-scale suppressed effects, although the interpretation may be subtler than in the quantum regime.

\section{Effective dynamics in the interior of the Schwarzschild black hole}
\label{sec:Effective dynamics in the interior of Schwarzschild black hole}
In Sec. \ref{sec:Non-deformed dynamics of black hole interior}, it has been shown that the choice of a new lapse function \eqref{eq:lapsNT} in the Hamiltonian \eqref{eq:H-class-N} yields classical solutions that recover the Schwarzschild metric \eqref{eq:sch-inter}. Now, we want to understand how the dynamics inside the black hole are modified when considering the deformed algebras \eqref{eqn:b_pb}-\eqref{eqn:c_pc}. It is important to note that the approach we will follow throughout the remainder of the paper is semiclassical.

The effective equations of motion are derived from the deformed algebra \eqref{eqn:b_pb}-\eqref{eqn:c_pc} and the Hamiltonian \eqref{eq:H-class-1}. This yields
\begin{align}
\frac{db}{dT}= & \left\{ b,H\right\} =-\frac{1}{2\gamma}\left(b^2+\gamma^{2}\right)\left(1+\beta_{b}b^{2}\right),\label{eq:EoM-diff-b-1}\\
\frac{dp_{b}}{dT}= & \left\{ p_{b},H\right\} =\frac{1}{\gamma}\left(bp_{b}+cp_{c}\right)\left(1+\beta_{b}b^{2}\right),\label{eq:EoM-diff-pb-1}\\
\frac{dc}{dT}= & \left\{ c,H\right\} =-\frac{2}{\gamma}\left(1+\beta_{c}c^{2}\right)bc,\label{eq:EoM-diff-c-1}\\
\frac{dp_{c}}{dT}= & \left\{ b,H\right\} =\frac{2}{\gamma}\left(1+\beta_{c}c^{2}\right)bp_{c}.\label{eq:EoM-diff-pc-1}
\end{align}
These equations reduce to the undeformed case when $\beta_b, \beta_c \rightarrow 0$. It can be seen that \eqref{eq:EoM-diff-pb-1}-\eqref{eq:EoM-diff-pc-1} are coupled; hence, they cannot be solved individually. We will follow an alternative procedure, first, we solve \eqref{eq:EoM-diff-b-1}
\begin{equation}
\frac{\gamma \sqrt{\beta_b} \tan^{-1} \left(\sqrt{\beta_b} b\right)-\tan^{-1} \left(\frac{b}{\gamma}\right)}{\gamma\left(1-\beta_b \gamma^2\right)}=\frac{T}{2 \gamma},
\end{equation}
It is not possible to directly solve for $b(T)$ from this expression, this procedure can be carried out numerically. However, to have analytical control over the solutions, we will undertake a perturbative approach through the deformation parameters $\beta_b$ and $\beta_c$. Then, we will consider the first-order approximation in the deformation parameter $\beta_b$, neglecting corrections of higher order $\mathcal{O}(\beta_b^2)$. In this way, the solution for $b(T)$ is obtained \footnote{A similar result was obtained numerically, confirming the approach taken.}
\begin{equation}
\label{eq: b-eff-T}
    b(T)]= - \gamma\tan\left( \frac{T}{2} \right) + \frac{\gamma^{3}\beta_{b}}{2}\left[ T\sec^{2}\left( \frac{T}{2} \right) - 2\tan\left( \frac{T}{2} \right)\right].
\end{equation}
We do not obtain the solution for $p_b(T)$ from \eqref{eq:EoM-diff-pb-1}. Instead, we will use the constraint \eqref{eq:weak-van} to solve for it
\begin{equation}
    p_b=-\frac{2bp_cc}{b^2+\gamma^2},
\end{equation}
taking into account that
\begin{equation}
\label{eq:cp_c_constat}
    cp_{c} = \text{constant} = C_{1},
\end{equation}
it is easy to see from \eqref{eq:EoM-diff-c-1} and \eqref{eq:EoM-diff-pc-1}, with this and considering \eqref{eq: b-eff-T}, the solution for $p_b(T)$ is obtained up to the first-order correction in $\beta_b$
\begin{equation}
    \label{eq: p_{b}(T)}
    p_{b}(T)=\frac{C_{1}}{\gamma}\sin{(T)}\left[ 1 +\beta_{b}\gamma^{2}\cos{(T)}\left( 1 - T\csc{(T)} \right) \right].
\end{equation}
The solution for $p_c(T)$ is obtained by integrating \eqref{eq:EoM-diff-pc-1} with the replacement of $c = C_1/p_c$ from \eqref{eq:cp_c_constat}. Thus,
\begin{multline}
p_{c}(T) = \pm e^{C_{2}/2}\cos^{4}\left( \frac{T}{2} \right)\Biggl( 1 +4\beta_{b}\gamma^{2}\left\lbrack T\tan\left( \frac{T}{2} \right) + 4\ln{\cos\left( \frac{T}{2} \right)} \right\rbrack\  -\\ 
\frac{\beta_{c}C_{1}^{2}}{e^{C_{2}}\cos^{8}\left( T/2 \right)}\Biggr)^{1/2},\label{eq: p_c(T)}
\end{multline}
where $C_2$ is an integration constant. For $c(T)$ is obtained from \eqref{eq:cp_c_constat}, as the inverse of \eqref{eq: p_c(T)}.

The values of the constants $C_1$ and $C_2$ are obtained by considering the classical limit of the effective solutions, i.e., when $\beta_b, \beta_c\rightarrow0$. Thus, from \eqref{eq:c_usually} and \eqref{eq:p_c_usually}, it can be seen that $p_c c=\mp\gamma G M L_0=C_1$, and also that $C_2=2\ln{(4G^2M^2)}$.

As done in the undeformed case, we need to express the effective solutions in terms of the Schwarzschild time $t$ instead of the $T$ parameter. For this purpose, we use the transformation proposed in Sec. \ref{sec:Non-deformed dynamics of black hole interior} of this work, where the relationship between $t$ and $T$ is given by $t=2GM\cos^2(T/2)$. So, \eqref{eq: b-eff-T}, \eqref{eq: p_{b}(T)}, and \eqref{eq: p_c(T)} become in
\begin{multline}
    \label{b-eff-t}
    b\left(t\right)= \pm\gamma\sqrt{\frac{2GM}{t}-1}+\frac{\beta_b\gamma^3}{2}\left[-2\cos^{-1}{\left(\pm\sqrt{\frac{t}{2GM}}\right)}\frac{2GM}{t}\pm\right.\\
    \left.2\sqrt{\frac{2GM}{t}-1}\right],
\end{multline}
\begin{multline}
    p_{b}\left(t\right)= \frac{l_c\sqrt{t(2GM-t)}}{\sqrt{-\beta_c}}\Biggl[1+\beta_b\gamma^2\left(\frac{t}{GM}-1\right)\times\\
    \left(1\mp2\cos^{-1}{\left(\pm\sqrt{\frac{t}{2GM}}\right)}\frac{GM}{\sqrt{t(2GM-t)}}\right)\Biggr], \label{pb(t)-eff}
\end{multline}
\begin{multline}
    p_{c}\left(t\right)=\pm t^2\Biggl(1+4\beta_b\gamma^2\Biggl[\ln{\left(\frac{t}{2GM}\right)^2}\pm2\cos^{-1}{\left(\pm\sqrt{\frac{t}{2GM}}\right)}\sqrt{\frac{2GM}{t}-1}\Biggr]\\
    +\frac{l_c^2\gamma^2G^2M^2}{t^4}\Biggr)^{1/2}.\label{pc(t)-eff}
\end{multline}
Here, we have removed the dependence on the fiducial length $L_0$ in our solutions by introducing a fundamental physical minimum length scale of our model, denoted as $l_c^2=-\beta_cL_0^2$, similar to what was done in \cite{Bosso}, where $l_c\propto l_{Pl}$ ( $l_{Pl}$ is the Planck length). A noteworthy result arises from equation \eqref{pc(t)-eff} when $t\to0$; the effective solution of $p_c$ reduces to
\begin{equation}
p_c(0)=\gamma GM l_{c}.\label{eq:pc_minimum}
\end{equation}
In this context, $p_c$ represents the square of the radius of the infalling two-spheres within the black hole's interior. Equation \eqref{eq:pc_minimum} demonstrates that $p_c$ maintains a non-zero value and does not diminish at any point within the black hole. A similar result was presented in \cite{Bosso}, the difference with ours lies in the choice of the Hamiltonian, whereas in \cite{Bosso} they opted for \eqref{eq:Lapse_funct_corichi_bosso} while we chose to work with \eqref{eq:lapsNT}.

\begin{figure}[htb!] 
\begin{center}
\includegraphics[scale=0.5]{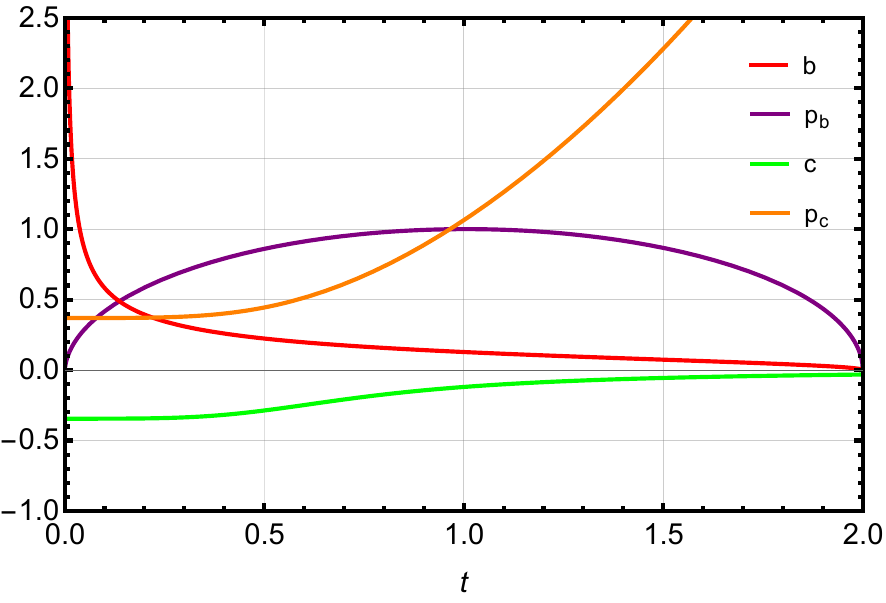} 
\caption{Here is presented the behavior of effective variables as a function of Schwarzschild time $t$ using the arbitrary values $\gamma= \ln{2}/\pi\sqrt{3}$, $G=1$ $M = 1$, $\beta_b=-0.6$, $\beta_c=-8.4$ and $l_c=2.9$. We have chosen the positive sign for $b(t)$ and the negative sign for $c(t)$ in their respective solutions.}
\label{fig:Plot_effect_solutions}
\end{center}
\end{figure}

In Figure \ref{fig:Plot_effect_solutions}, we graphically depict the behavior of the effective solutions \eqref{b-eff-t}-\eqref{pb(t)-eff}, \eqref{pc(t)-eff}-\eqref{eq:cp_c_constat}. From this Figure, it can be seen that the behavior of the effective solutions differs from that of the classical ones, presented in Figure \ref{fig1}. These differences are more noticeable in the behavior of $c$ and $p_c$. The selection of the negative sign of $\beta_c$ is made to ensure that the fundamental length $l_c$ is real, and therefore \eqref{eq:pc_minimum}.

If we consider that the effective corrections introduced by $\beta_c$ are small at the semiclassical level, similar to $\beta_b$, then we can focus solely on the first-order approximations in $\beta_c$ and disregard higher-order corrections $\mathcal{O}(\beta_c^2)$. In this sense, we can expand \eqref{pc(t)-eff} in a Taylor series. Knowing that  $(1+x)^{1/2}\sim1+x/2+\cdots$, one can conclude that
 \begin{multline}
      p_{c}\left(t\right)=\pm t^2\Biggl(1+2\beta_b\gamma^2\Biggl[\ln{\left(\frac{t}{2GM}\right)^2}\pm2\cos^{-1}{\left(\pm\sqrt{\frac{t}{2GM}}\right)}\sqrt{\frac{2GM}{t}-1}\Biggr]+\\
      \frac{l_c^2\gamma^2G^2M^2}{2t^4}\Biggr),\label{pc(t)-eff_expand}
 \end{multline}
 and, based on \eqref{eq:cp_c_constat}, the solution for $c(t)$ is obtained
 \begin{multline}
      c\left(t\right)=\mp\frac{\gamma G M l_c}{\sqrt{-\beta_c}t^2} \Biggl(1-2\beta_b\gamma^2\Biggl[\ln{\left(\frac{t}{2GM}\right)^2}\pm\\
      2\cos^{-1}{\left(\pm\sqrt{\frac{t}{2GM}}\right)}\sqrt{\frac{2GM}{t}-1}\Biggr]-\frac{l_c^2\gamma^2G^2M^2}{2t^4}\Biggr).\label{c(t)-eff_expand}
 \end{multline}
The dynamic of a particle inside the black hole is modified by these solutions, and this modification arises due to the presence of the deformation parameters $\beta_b$ and $\beta_c$. In the non-effective limit, where $\beta_b,\beta_c\rightarrow0$, the effective solutions are reduced to the classical ones.

\begin{figure}[htb!] 
\begin{center}
\includegraphics[scale=0.55]{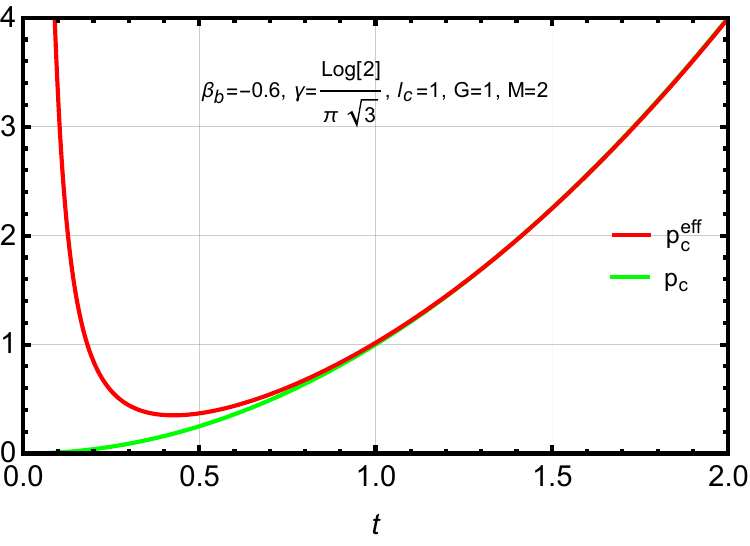} 
\caption{Comparison between the classical and effective solution of $p_c$. In the quantum region, the effective solution experiences a bounce towards large $p_c$ values without reaching the $t=0$ region, whereas the classical solution converges to $p_c=0$ in that region.}
\label{fig:Plot_pc_eff_compar}
\end{center}
\end{figure}

Figure \ref{fig:Plot_pc_eff_compar} shows the comparison between the classical solution of $p_c$, \eqref{eq:p_c_usually}, and the effective solution given in \eqref{pc(t)-eff_expand}, in the range $t\ensuremath{\in}[0,2]$. As we approach the $t=0$ region, it can be observed that $p_c^{\text{eff}}$ reaches a minimum value, as seen in Figure \ref{fig:Plot_pc_eff_compar}, and then grows without reaching $p_c=0$, whereas the classical solution converges to $p_c=0$ in that region. This behavior can be interpreted as a bounce, which is a result of the effective dynamics we have considered.

\begin{figure}[htb!] 
\begin{center}
\includegraphics[scale=0.58]{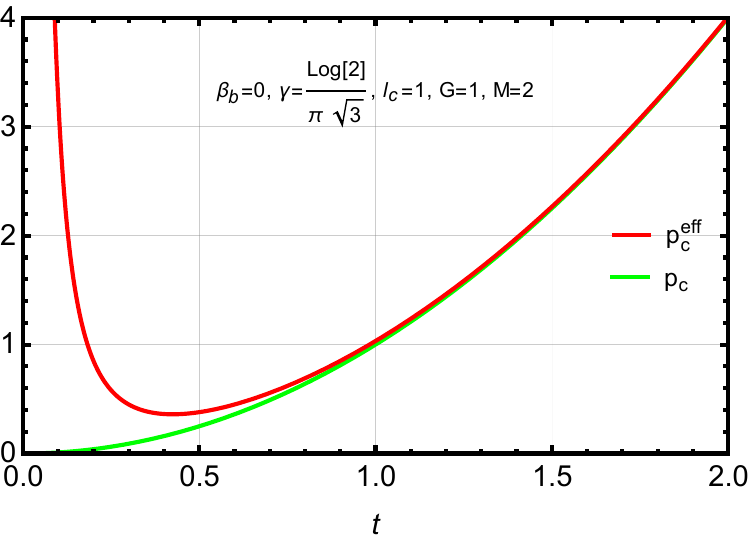} 
\caption{Graphical representation of equation \eqref{pc(t)-eff_expand} for $\beta_b=0$. It is clearly seen that the effective parameter $\beta_b$ does not affect the existence of the bounce in $p_c$.}
\label{fig:Plot_pc_eff_bb_0}
\end{center}
\end{figure}

It is interesting to note that the effective parameter $\beta_b$ does not affect the presence of the bounce in \eqref{pc(t)-eff_expand}, as seen in Figure \ref{fig:Plot_pc_eff_bb_0} where $\beta_b=0$ was fixed. Therefore, only the presence of the fundamental length $l_c$ (or the effective parameter $\beta_c$) allows the existence of the bounce in $p_c$.

\subsection{Minimum values of $p_c^{\text{eff}}$ and $p_b^{\text{eff}}$}
We want to determine the minimum value of $p_c^{\text{eff}}$ from \eqref{pc(t)-eff_expand}. The minimum value of $p_c$ is found at $t_{\text{min}}$ (see Figure \ref{fig:Plot_pc_eff_compar})
\begin{equation}
p_c(t_{\text{min}})=p_{c}^{\textrm{min}},\label{eq:pc_minimum_eff}
\end{equation}
where $t_{\text{min}}$ depends on the values that the parameters $M$ and $l_c$ can take, fixing $G$, $\gamma$, and $\beta_b$. Table \ref{tab:1} shows the different values that $t_{\text{min}}$, and hence $p_{c}^{\textrm{min}}$, can take for different values of $M$, keeping $l_c$ fixed. On the other hand, in Table \ref{tab:2}, the value of $M$ is kept fixed while $l_c$ varies, clearly showing the dependence of $t_{\text{min}}$ on these parameters. Furthermore, we can verify that $p_c$ does not vanish inside the black hole; $p_c=0$ only occurs when the effective corrections vanish, i.e., when $\beta_b,\ \beta_c\to0$ (see Table \ref{tab:2}).
\begin{table}[h]
\caption{Different values of $t_{\text{min}}$ and  $p_{c}^{\textrm{min}}$ for several $M$, keeping $G=1$, $\gamma=\ln{2}/\pi\sqrt{3}$, $\beta_b=-0.5$, and $l_c =1$ fixed.}
\label{tab:1}
    \begin{tabularx}{\linewidth}{@{} lXl @{}}
\toprule
 $M$ &$t_{\text{min}}$&$p_{c}^{\text{min}}$\\
\midrule
 1 & 0.30  & 0.18 \\
 3 & 0.52  & 0.53 \\
 5 & 0.68  & 0.87   \\
 8 & 0.86  & 1.38  \\
\bottomrule
\end{tabularx}
\end{table}
\begin{table}[h]
\caption{Values of $t_{\text{min}}$ and $p_{c}^{\textrm{min}}$ for different $l_c$, keeping $G=1$, $\gamma=\ln{2}/\pi\sqrt{3}$, $\beta_b=0$, and $M =2$ fixed.}
\label{tab:2}
    \begin{tabularx}{\linewidth}{@{} lXl @{}}
\toprule
$l_c$ &$t_{\text{min}}$&$p_{c}^{\text{min}}$\\
\midrule
 0   & 0     & 0 \\
 0.5 & 0.30  & 0.18 \\
 0.8 & 0.38  & 0.29 \\
 1   & 0.42  & 0.36 \\
\bottomrule
\end{tabularx}
\end{table}

Given that $l_c\propto l_{Pl}=\sqrt{\hbar G/c^3}$, we observe that as $\hbar G\to0$, $l_c\to0$, and therefore $p_c\to0$. It can be interpreted that the bounce in $p_c$ has a quantum gravitational origin \cite{Corichi:2015xia}, despite the semiclassical approach used. Based on the consistent limits of GUP discussed in \cite{Bosso_2023}, we can assert that the corrections to classical dynamics introduced by $\beta_b$ and $\beta_c$ are consequences of Planck-scale suppressed effects.

On the other hand, the minimum value of $p_b$ is found at $t=0$, this can be calculated from \eqref{pb(t)-eff}
\begin{equation}
    p_b^{\text{min}}=-\frac{\pi l_c\beta_b \gamma ^2 G M}{\sqrt{-\beta_c}},\label{eq:pb_minimum}
\end{equation}
which differs from the value obtained from \eqref{eq:b_pb_usually} at $t\rightarrow0$. This result has not been reported before in the literature. In \eqref{eq:pb_pc_physical_area} we saw that $p_b$ represents a physical area $A_{x,\theta}=A_{x,\phi}=2\pi p_b$, so this quantity must be real and positive. This leads us to conclude that the deformation parameters must satisfy $\beta_b,\beta_c<0$. The deformation parameters can take both positive and negative values \cite{BIZET2023137636}, depending on the physical situation being considered; both possibilities are admissible.

From \eqref{eq:pc_minimum_eff} and \eqref{eq:pb_minimum}, we can see that the minimum values of $p_c$ and $p_b$ are affected by both the Barbero-Immirzi parameter $\gamma$ and the deformation parameters $\beta_b$ and $\beta_c$, which are effective, as well as the gravitational constant $G$ and the mass of the black hole $M$, which are gravitational quantities. This implies that the presence of these minimum quantities is due to an effective-gravitational effects. In the limit $\beta_b, \beta_c \rightarrow 0$, these effects vanish, and our expressions reduce to the classical results.

\subsection{Effective correction to the event horizon}
A modification occurs at the black hole horizon. In this region, $p_c$ in \eqref{pc(t)-eff_expand} takes the value
\begin{equation}
    p_c^{\text{hor}}=4 G^2 M^2+\frac{\gamma ^2l_c^2}{8}.\label{eq:pc_effec_horizon}
\end{equation}
On the contrary, in \eqref{pb(t)-eff} it can be seen that at the horizon $p_b(2GM)=0$, just like in the non-deformed case \eqref{eq:b_pb_usually}. From \eqref{eq:pb_pc_physical_area}, it can be seen that the physical area $A_{\theta\phi}=\pi p_c$ \cite{Chiou:2008nm}, that in the horizon represents the black hole area, is modified due to \eqref{eq:pc_effec_horizon} by
\begin{equation}
   4 A_{\theta\phi}=16\pi G^2 M^2+\frac{\pi\gamma ^2l_c^2}{2}.\label{eq:Mod_area_horizon}
\end{equation}
The first term on the right-hand side of this expression is the usual area of the black hole, while the second term is an effective correction because it only depends on the parameters $\gamma$ and $l_c$. If the mass of the black hole is large, the first term in this expression dominates, making the semiclassical correction negligible in this classical regime. On the other hand, the area $A_{x\theta}=A_{x\phi}=2\pi p_b$ \cite{Chiou:2008nm} is zero at the horizon, due to $p_b(2GM)=0$, implying that there is no correction term and it is not proportional to $M$. In \cite{Bosso}, the correction to the effective $p_b^{\text{eff}}(2GM)$ at the horizon is proportional to $M$, which means it is larger for more massive black holes. In our effective model, with a different lapse function, there is no such correction, and it will be zero for any mass $M$ of the black hole.

\subsection{Minimum area and the emergence of a new horizon}
Because $p_c$ reaches a minimum value as we approach the $t=0$ region, see \eqref{eq:pc_minimum_eff}, we can define a minimum physical area
\begin{equation}
    A_{\theta\phi}^{\text{min}}=\pi p_{c}^{\textrm{min}}.\label{eq:Minim_area}
\end{equation}
This quantity can be interpreted as the area of an event horizon, whose value depends not only on the gravitational parameters $G$ and $M$ but also on the parameters $\gamma$ and $l_c$. Thus, we conclude that, due to the introduced effective dynamics, a new event horizon arises.

\subsection{Singularity resolution}
We will analyze the Kretschmann scalar in three regions of interest: the horizon, the effective region (where $p_c$ reaches a minimum value), and in the classical singularity $t=0$.

The Kretschmann scalar is corrected at the black hole horizon due to the discussed effective dynamics. In this region, from \eqref{eq:pc_effec_horizon}, it takes the form
\begin{equation}
    K(p_c^{\text{hor}})=\frac{3}{4G^4M^4}\left(1-\frac{3}{32}\frac{\gamma^2l_c^2}{G^2M^2}\right).
\end{equation}
From this expression, we can see that the correction term inversely depends on the mass of the black hole and directly depends on the parameters $\gamma$ and $l_c$. Therefore, as $M$ increases, the correction term becomes negligible in this limit. Modifications to the horizon have already been made in the literature; for example, we can cite references \cite{Chiou:2008nm, Bosso, Sobrinho_2023}. These modifications occur in the context of LQG.

As we approach the $t=0$ region, the effective solution for $p_c$ attains a minimum value in $t_{\text{min}}$ (see Figure \ref{fig:Plot_pc_eff_compar}), \eqref{eq:pc_minimum_eff}. Here, the Kretschmann scalar takes the value
\begin{equation}
    K(p_{c}^{\textrm{min}})=\frac{48 G^2 M^2}{(p_{c}^{\textrm{min}})^3},\label{eq:Kretschmann_scalar_singularity}
\end{equation}
which is finite value governed by $\gamma$, $l_c$ parameters, and the gravitational ones, $G$ and $M$. For this reason, the existence of a new horizon is proposed, whose radius is given by \eqref{eq:pc_minimum_eff}.

After passing through the effective region characterized by $t_{\text{min}}$, the solution \eqref{pc(t)-eff_expand} of $p_c$ undergoes a rebound towards very large values. Strictly speaking, as $t\rightarrow0$, this implies that $p_c\rightarrow\infty$, and as a consequence, the Kretschmann scalar becomes zero in those regions, indicating that spacetime is flat

\begin{equation}
    K(p_c\to\infty)\to0,\quad\quad \text{when}\quad\quad t\to 0.\label{eq:k_0_t_0_deform}
\end{equation}
This does not happen in the classical case, where $K\to\infty$ when $t\to0$.

In conclusion, the classical singularity of the black hole is replaced by a bounce that connects the interior of the black hole to the interior of a white hole, and this bounce occurs in the Planck scale, because $l_c\propto l_{pl}$. This result is not new; a similar interpretation was presented in \cite{Modesto, Chiou:2008nm, Corichi} within the context of LQG. This demonstrates an interesting connection between the formalism inspired by the minimal uncertainty approach developed in this work and that of LQG.

\subsection{Effective components of the Kantowski-Sachs metric}
Considering \eqref{pc(t)-eff_expand} and the lapse function in \eqref{eq:lapsNT}, the effective component $g_{00}$ of the Kantowski-Sachs metric take the form
\begin{multline}
    N_{\text{eff}}^2(t)=\frac{t^2}{(16\pi G)^2}\Biggl(1+2\beta_b\gamma^2\Biggl[\ln{\left(\frac{t}{2GM}\right)^2}+\\
      2\cos^{-1}{\left(\sqrt{\frac{t}{2GM}}\right)}\sqrt{\frac{2GM}{t}-1}\Biggr]+\frac{l_c^2\gamma^2G^2M^2}{2t^4}\Biggr),\label{eq:Effect_lapse_funct_t}
\end{multline}
and, from \eqref{pb(t)-eff}, we have
\begin{equation}
     g_{xx}^{\text{eff}}=-\frac{\beta_c(p_b^{\text{eff}}(t))^2}{l_c^2 |p_c^{\text{eff}}(t)|},\label{eq:gxx_effect_compont_metric}
\end{equation}
where $p_b^{\text{eff}}(t)$ and $p_c^{\text{eff}}(t)$ represents the effective solutions in \eqref{pb(t)-eff} and \eqref{pc(t)-eff_expand}, respectively. Here we choose $\text{sign}(p_c)=+1$. The angular component of the metric is given by $g_{\Omega\Omega}=p_c$, so we do not need to express its effective form explicitly here.

\begin{figure}[htb!] 
\begin{center}
\includegraphics[scale=0.6]{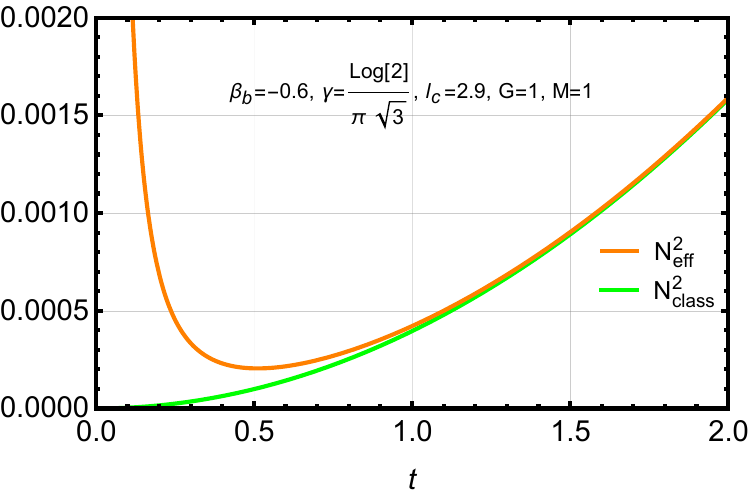} 
\caption{Graphical representation of the classical and effective lapse functions, $N_{\text{class}}$ and $N_{\text{eff}}$ respectively.}
\label{fig:Plot_N_compari}
\end{center}
\end{figure}

In Figure \ref{fig:Plot_N_compari}, a graphical comparison is made between the classical lapse function $N^2_{\text{class}}(t)$ and the one obtained from effective dynamics, $N^2_{\text{eff}}(t)$, in \eqref{eq:Effect_lapse_funct_t}. From this graph, it can be observed that both lapse functions are identical as we approach the event horizon ($t=2GM$), but they differ completely as we move toward the $t=0$ region. In the semiclassical region governed by the parameters $\gamma$ and $l_c$, $N^2_{\text{eff}}(t)$ acquires a minimum value before rebounding towards larger values, whereas $N^2_{\text{class}}$ remains unaffected and tends to zero as $t$ approaches zero. A similar behavior was described in \cite{Modesto} where modifications to the Schwarzschild solution were calculated using a semiclassical analysis of loop quantum black holes. In this approach, the author \cite{Modesto} finds a metric inside the event horizon that coincides with the Schwarzschild solution near the horizon but deviates substantially at the Planck scale. Particularly, he obtains a bounce of the $S^2$ sphere for a minimum value of the radius, and it is possible to have another event horizon close to the $r=0$ point.

\begin{figure}[htb!] 
\begin{center}
\includegraphics[scale=0.6]{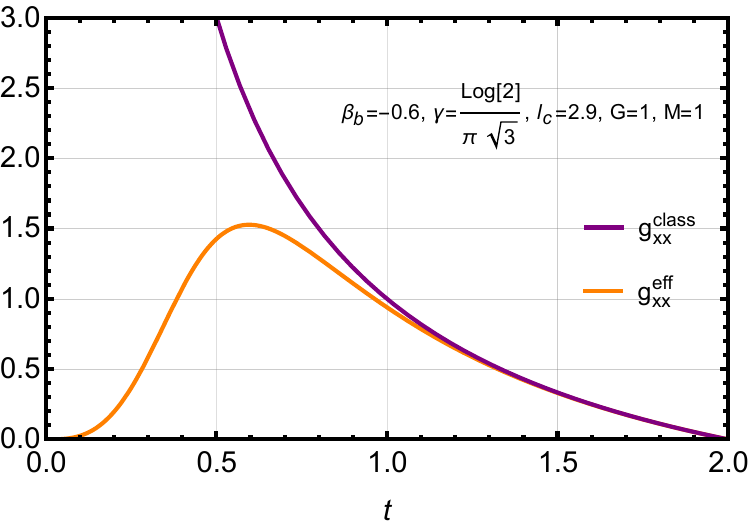} 
\caption{The metric components $g_{xx}$ are compared for the classical and effective cases. The purple curve represents the classical component, while the orange one represents the effective component.}
\label{fig:Plot_g_xx_compari}
\end{center}
\end{figure}

Figure \ref{fig:Plot_g_xx_compari} presents the graph of $g_{xx}^{\text{class}}$ and $g_{xx}^{\text{eff}}$ as functions of $t$. In this case, as in the case of Figure \ref{fig:Plot_N_compari}, it can be observed that the effective metric component reduces to the classical metric as we approach the event horizon, but they differ in the Planck region, governed by the fundamental length scale $l_c$ and the Barbero-Immirzi parameter $\gamma$. This result is analogous to what is found in \cite{Modesto} in the context of LQG.

We can interpret this behavior as follows: As we approach the Planck region, the effective component $g_{xx}^{\text{eff}}$ of the metric reaches a maximum in this region and decays to zero at $t=0$, which does not occur with $g_{xx}^{\text{class}}$, which grows to very large values as $t\rightarrow0$. On the other hand, as we have seen from Figure \ref{fig:Plot_N_compari}, the component $g_{tt}^{\text{eff}}$ undergoes a rebound towards large values in the Planck region, while $g_{tt}^{\text{class}}$ converges to zero at $t\rightarrow0$. This can be seen as a change in the nature of the interval, from space-like (inside the black hole) to time-like (outside the white hole), due to the bounce.

\section{Conclusions and discussions}
\label{sec:Conclusions}
In this work, we consider the Hamiltonian derived from Loop Quantum Geometry (LQG) to describe the interior of a Schwarzschild black hole in terms of Ashtekar-Barbero variables. However, instead of using any of the lapse functions considered in the literature \cite{Ashtekar, Chiou:2008nm}, we propose a different one, as shown in \eqref{eq:lapsNT}, with the aim of simplifying the model's Hamiltonian for quantization \eqref{eq:H-class-1}.

From this Hamiltonian, we obtain the equations of motion, which, when solved, recover the classical metric solutions for an appropriate transformation \eqref{eq:Transf_T_to_t}. This indicates that our choice of the new lapse function, \eqref{eq:lapsNT}, is reasonable.

Using the modification in the Poisson algebra proposed in \cite{Bosso}, \eqref{eqn:b_pb} and \eqref{eqn:c_pc}, and the new Hamiltonian \eqref{eq:H-class-1}, we obtain the effective equations of motion \eqref{eq:EoM-diff-b-1}-\eqref{eq:EoM-diff-pc-1}. The solutions to these effective equations are obtained by considering only the first-order corrections in the deformation parameters $\beta_b$ and $\beta_c$, \eqref{b-eff-t}, \eqref{pb(t)-eff}, \eqref{pc(t)-eff_expand} and \eqref{c(t)-eff_expand}. From these solutions, the dependence on the fiducial length $L_0$ is removed by introducing a fundamental length $l_c\propto l_{Pl}$ (where $l_{Pl}$ is the Planck length), similar to what was done in \cite{Bosso}, with the aim of ensuring that scalar quantities (such as the Kretschmann scalar) do not depend on the chosen length scale.

An intriguing finding, not documented in prior literature, emerges when $\beta_b<0$ is selected in \eqref{pb(t)-eff}. This outcome reveals the existence of a $p_b^{\text{min}}$ (refer to \eqref{eq:pb_minimum}) at the classical singularity $t=0$. It is characterized in terms of the fundamental length $l_c$, the Barbero-Immirzi parameter $\gamma$, suggesting its quantum origin that diminishes in the classical limit ($\hbar G\to 0$), along with the gravitational parameters $G$ and $M$. Interestingly, on the $p_b$ horizon, it remains impervious to effective corrections, maintaining $p_b^{\text{eff}}(2GM)=0$, mirroring the behavior observed in the classical scenario.

As a consequence of the effective corrections introduced by $\beta_c$ in the dynamics, $p_c$ acquires a minimum value, given in \eqref{eq:pc_minimum_eff} for $\beta_c<0$, which is different from zero as it approaches the classical singularity at $t=0$. This effect does not occur in the non-effective case, where $p_c\to0$ as $t\to0$. In \cite{Bosso}, starting from a different Hamiltonian than the one considered in this work, the condition $\beta_c<0$ leads to a minimum radius on the 2-sphere, but not to a bounce. In our case, choosing $\beta_c < 0$ leads to a bounce, which originates from semiclassical corrections and vanishes in the classical limit, that connects the interior of a black hole with the interior of a white hole (see Figure \ref{fig:Plot_pc_eff_compar}). The event horizon of the black hole undergoes a modification, \eqref{eq:Mod_area_horizon}, which is negligible compared to the mass of the black hole. On the other hand, the event horizon of the white hole has an effective origin, \eqref{eq:Minim_area}, through which two universes are connected. The bounce occurs on the Planck scale because $l_c\propto l_{Pl}$. Additionally, from Figure \ref{fig:Plot_pc_eff_compar}, we can see that $p_c$, after the bounce, grows indefinitely. This implies that the Kretschmann scalar is zero in that region, a sign that the spacetime outside the white hole is flat (see \eqref{eq:k_0_t_0_deform}).

It is noteworthy that this outcome has been previously discussed by other authors within the framework of Loop Quantum Gravity (LQG) \cite{Modesto, Chiou:2008nm, Corichi}. In the context of polymer quantization employed in LQG, a parameter known as the polymer scale is introduced, establishing a minimum scale for the model and altering the algebra of the theory at the quantum level. This can be viewed as an effective modification of the classical algebra. Hence, we interpret that there exists an equivalence between the formalism developed in this work and that of LQG, a correlation we leave for future investigation.

\acknowledgments

We thank Prof. O. Obreg\'on for the recommendations and suggestions made. Additionally, we express our gratitude to J. C. Jim\'enez for their careful and critical reading. BM thanks the support of UNSA-INVESTIGA by PTTMD-41-2023-UNSA Project. WY thanks the University of Guanajuato for the support through the grants CIIC 168/2023 "Non-extensive Entropies Independent of Parameters", CIIC 251/2024 "Teor\'ias efectivas de cuerdas y exploraciones de aprendizaje de m\'aquina", and CIIC 156/2024 "Generalized Uncertainty Principle, Non-extensive Entropies, and General Relativity", as well as the CONAHCyT Grant CBF2023-2024-2923 "Implicaciones del Principio de Incertidumbre Generalizado (GUP) en cosmolog\'ia cu\'antica, gravitaci\'on y su conexi\'on con entrop\'ias no-extensivas".




\bibliographystyle{JHEP}

\bibliography{biblio}

\providecommand{\href}[2]{#2}\begingroup\raggedright\begin{thebibliography}{10}

\bibitem{Abhay_Ashtekar_2004}
A.~Ashtekar and J.~Lewandowski, \emph{Background independent quantum gravity: a status report}, \href{https://doi.org/10.1088/0264-9381/21/15/R01}{\emph{Classical and Quantum Gravity} {\bfseries 21} (2004) R53}.

\bibitem{Thiemann2007}
T.~Thiemann, \emph{Loop quantum gravity: An inside view},  in \emph{Approaches to Fundamental Physics: An Assessment of Current Theoretical Ideas}, I.-O.~Stamatescu and E.~Seiler, eds., (Berlin, Heidelberg), pp.~185--263, Springer Berlin Heidelberg (2007), \href{https://doi.org/https://doi.org/10.1007/978-3-540-71117-9_10}{DOI}.

\bibitem{Thiemann2003}
T.~Thiemann, \emph{Lectures on loop quantum gravity},  in \emph{Quantum Gravity: From Theory to Experimental Search}, D.J.W.~Giulini, C.~Kiefer and C.~L{\"a}mmerzahl, eds., (Berlin, Heidelberg), pp.~41--135, Springer Berlin Heidelberg (2003), \href{https://doi.org/10.1007/978-3-540-45230-0_3}{DOI}.

\bibitem{Corichi:2015xia}
A.~Corichi and P.~Singh, \emph{{Loop quantization of the Schwarzschild interior revisited}}, \href{https://doi.org/10.1088/0264-9381/33/5/055006}{\emph{Class. Quant. Grav.} {\bfseries 33} (2016) 055006} [\href{https://arxiv.org/abs/1506.08015}{{\ttfamily 1506.08015}}].

\bibitem{Chiou2008}
D.-W.~Chiou, \emph{Phenomenological dynamics of loop quantum cosmology in kantowski-sachs spacetime}, \href{https://doi.org/10.1103/PhysRevD.78.044019}{\emph{Phys. Rev. D} {\bfseries 78} (2008) 044019}.

\bibitem{Bosso}
P.~Bosso, O.~Obreg\'on, S.~Rastgoo and W.~Yupanqui, \emph{Deformed algebra and the effective dynamics of the interior of black holes}, \href{https://doi.org/10.1088/1361-6382/ac025f}{\emph{Class. Quantum Grav.} {\bfseries 38} (2021) 145006}.

\bibitem{Abhay_Ashtekar_2003}
A.~Ashtekar, S.~Fairhurst and J.L.~Willis, \emph{Quantum gravity, shadow states and quantum mechanics}, \href{https://doi.org/10.1088/0264-9381/20/6/302}{\emph{Classical and Quantum Gravity} {\bfseries 20} (2003) 1031}.

\bibitem{PhysRevD.76.044016}
A.~Corichi, T.~Vuka\ifmmode~\check{s}\else \v{s}\fi{}inac and J.A.~Zapata, \emph{Polymer quantum mechanics and its continuum limit}, \href{https://doi.org/10.1103/PhysRevD.76.044016}{\emph{Phys. Rev. D} {\bfseries 76} (2007) 044016}.

\bibitem{PhysRevD.95.065026}
H.A.~Morales-T\'ecotl, S.~Rastgoo and J.C.~Ruelas, \emph{Path integral polymer propagator of relativistic and nonrelativistic particles}, \href{https://doi.org/10.1103/PhysRevD.95.065026}{\emph{Phys. Rev. D} {\bfseries 95} (2017) 065026}.

\bibitem{PhysRevD.92.104029}
H.A.~Morales-T\'ecotl, D.H.~Orozco-Borunda and S.~Rastgoo, \emph{Polymer quantization and the saddle point approximation of partition functions}, \href{https://doi.org/10.1103/PhysRevD.92.104029}{\emph{Phys. Rev. D} {\bfseries 92} (2015) 104029}.

\bibitem{FLORESGONZALEZ2013394}
E.~Flores-Gonz\'{a}lez, H.A.~Morales-T\'{e}cotl and J.D.~Reyes, \emph{Propagators in polymer quantum mechanics}, \href{https://doi.org/https://doi.org/10.1016/j.aop.2013.05.005}{\emph{Annals of Physics} {\bfseries 336} (2013) 394}.

\bibitem{PhysRevD.52.1108}
A.~Kempf, G.~Mangano and R.B.~Mann, \emph{Hilbert space representation of the minimal length uncertainty relation}, \href{https://doi.org/10.1103/PhysRevD.52.1108}{\emph{Phys. Rev. D} {\bfseries 52} (1995) 1108}.

\bibitem{G.Veneziano_1986}
G.~Veneziano, \emph{A stringy nature needs just two constants}, \href{https://doi.org/10.1209/0295-5075/2/3/006}{\emph{Europhysics Letters} {\bfseries 2} (1986) 199}.

\bibitem{SCARDIGLI199939}
F.~Scardigli, \emph{Generalized uncertainty principle in quantum gravity from micro-black hole gedanken experiment}, \href{https://doi.org/https://doi.org/10.1016/S0370-2693(99)00167-7}{\emph{Physics Letters B} {\bfseries 452} (1999) 39}.

\bibitem{BIZET2023137636}
N.C.~Bizet, O.~Obreg\'{o}n and W.~Yupanqui, \emph{Modified entropies as the origin of generalized uncertainty principles}, \href{https://doi.org/https://doi.org/10.1016/j.physletb.2022.137636}{\emph{Physics Letters B} {\bfseries 836} (2023) 137636}.

\bibitem{10.1088/1361-6382/ad4fd7}
P.~Bosso, O.J.~Obregon, S.~Rastgoo and W.~Yupanqui, \emph{Black hole interior quantization: a minimal uncertainty approach}, {\emph{Classical and Quantum Gravity} (2024) }.

\bibitem{Ashtekar}
A.~Ashtekar and M.~Bojowald, \emph{Quantum geometry and the schwarzschild singularity}, \href{https://doi.org/10.1088/0264-9381/23/2/008}{\emph{Class. Quantum Grav.} {\bfseries 23} (2006) 391}.

\bibitem{Barbero}
J.F.~Barbero~G., \emph{Real ashtekar variables for lorentzian signature space-times}, \href{https://doi.org/10.1103/PhysRevD.51.5507}{\emph{Phys. Rev. D} {\bfseries 51} (1995) 5507}.

\bibitem{Chiou:2008nm}
D.-W.~Chiou, \emph{{Phenomenological loop quantum geometry of the Schwarzschild black hole}}, \href{https://doi.org/10.1103/PhysRevD.78.064040}{\emph{Phys. Rev. D} {\bfseries 78} (2008) 064040} [\href{https://arxiv.org/abs/0807.0665}{{\ttfamily 0807.0665}}].

\bibitem{Modesto}
L.~Modesto, \emph{Black hole interior from loop quantum gravity}, \href{https://doi.org/10.1155/2008/459290}{\emph{Advances in High Energy Physics} {\bfseries 2008} (2008) 459290}.

\bibitem{universe8070349}
S.~Rastgoo and S.~Das, \emph{Probing the interior of the schwarzschild black hole using congruences: Lqg vs. gup}, \href{https://doi.org/10.3390/universe8070349}{\emph{Universe} {\bfseries 8} (2022) }.

\bibitem{kennard1927quantenmechanik}
E.H.~Kennard, \emph{Zur quantenmechanik einfacher bewegungstypen}, {\emph{Zeitschrift f{\"u}r Physik} {\bfseries 44} (1927) 326}.

\bibitem{guth1929gruppentheorie}
E.~Guth, \emph{Gruppentheorie und quantenmechanik: H. weyl s. hirzel, leipzig 1928 viii+ 288 seiten, rm 20}, {\emph{Monatshefte f{\"u}r Mathematik und Physik} {\bfseries 36} (1929) A48}.

\bibitem{PhysRev.34.163}
H.P.~Robertson, \emph{The uncertainty principle}, \href{https://doi.org/10.1103/PhysRev.34.163}{\emph{Phys. Rev.} {\bfseries 34} (1929) 163}.

\bibitem{G_Veneziano_1986}
G.~Veneziano, \emph{A stringy nature needs just two constants}, \href{https://doi.org/10.1209/0295-5075/2/3/006}{\emph{Europhysics Letters} {\bfseries 2} (1986) 199}.

\bibitem{MAGGIORE199365}
M.~Maggiore, \emph{A generalized uncertainty principle in quantum gravity}, \href{https://doi.org/https://doi.org/10.1016/0370-2693(93)91401-8}{\emph{Physics Letters B} {\bfseries 304} (1993) 65}.

\bibitem{Bosso_2023}
P.~Bosso, G.G.~Luciano, L.~Petruzziello and F.~Wagner, \emph{30 years in: Quo vadis generalized uncertainty principle?}, \href{https://doi.org/10.1088/1361-6382/acf021}{\emph{Classical and Quantum Gravity} {\bfseries 40} (2023) 195014}.

\bibitem{PhysRevD.79.125007}
M.~Zarei and B.~Mirza, \emph{Minimal uncertainty in momentum: The effects of ir gravity on quantum mechanics}, \href{https://doi.org/10.1103/PhysRevD.79.125007}{\emph{Phys. Rev. D} {\bfseries 79} (2009) 125007}.

\bibitem{MUREIKA201988}
J.~Mureika, \emph{Extended uncertainty principle black holes}, \href{https://doi.org/https://doi.org/10.1016/j.physletb.2018.12.009}{\emph{Physics Letters B} {\bfseries 789} (2019) 88}.

\bibitem{doi:10.1142/S0217732319502043}
W.S.~Chung, H.~Hassanabadi and N.~Farahani, \emph{Klein–gordon oscillator in the presence of the minimal momentum}, \href{https://doi.org/10.1142/S0217732319502043}{\emph{Modern Physics Letters A} {\bfseries 34} (2019) 1950204} [\href{https://arxiv.org/abs/https://doi.org/10.1142/S0217732319502043}{{\ttfamily https://doi.org/10.1142/S0217732319502043}}].

\bibitem{ALI2009497}
A.F.~Ali, S.~Das and E.C.~Vagenas, \emph{Discreteness of space from the generalized uncertainty principle}, \href{https://doi.org/https://doi.org/10.1016/j.physletb.2009.06.061}{\emph{Physics Letters B} {\bfseries 678} (2009) 497}.

\bibitem{Bosso_2020}
P.~Bosso and O.~Obreg\'{o}n, \emph{Minimal length effects on quantum cosmology and quantum black hole models}, \href{https://doi.org/10.1088/1361-6382/ab6038}{\emph{Classical and Quantum Gravity} {\bfseries 37} (2020) 045003}.

\bibitem{PhysRevD.65.125028}
L.N.~Chang, D.~Minic, N.~Okamura and T.~Takeuchi, \emph{Effect of the minimal length uncertainty relation on the density of states and the cosmological constant problem}, \href{https://doi.org/10.1103/PhysRevD.65.125028}{\emph{Phys. Rev. D} {\bfseries 65} (2002) 125028}.

\bibitem{PhysRevD.66.026003}
S.~Benczik, L.N.~Chang, D.~Minic, N.~Okamura, S.~Rayyan and T.~Takeuchi, \emph{Short distance versus long distance physics: The classical limit of the minimal length uncertainty relation}, \href{https://doi.org/10.1103/PhysRevD.66.026003}{\emph{Phys. Rev. D} {\bfseries 66} (2002) 026003}.

\bibitem{CASADIO2020135558}
R.~Casadio and F.~Scardigli, \emph{Generalized uncertainty principle, classical mechanics, and general relativity}, \href{https://doi.org/https://doi.org/10.1016/j.physletb.2020.135558}{\emph{Physics Letters B} {\bfseries 807} (2020) 135558}.

\bibitem{Sobrinho_2023}
F.C.~Sobrinho, H.A.~Borges, I.P.R.~Baranov and S.~Carneiro, \emph{On the horizon area of effective loop quantum black holes}, \href{https://doi.org/10.1088/1361-6382/acdbff}{\emph{Classical and Quantum Gravity} {\bfseries 40} (2023) 145003}.

\bibitem{Corichi}
A.~Corichi and P.~Singh, \emph{Loop quantization of the schwarzschild interior revisited}, \href{https://doi.org/10.1088/0264-9381/33/5/055006}{\emph{Class. Quantum Grav.} {\bfseries 33} (2016) 055006}.

\end{thebibliography}\endgroup
\end{document}